\documentclass[fleqn,usenatbib]{mnras}

\usepackage{newtxtext,newtxmath}

\usepackage[T1]{fontenc}
\usepackage{ae,aecompl}

\usepackage{graphicx}	
\usepackage{amsmath}	
\usepackage[export]{adjustbox}
\usepackage{caption}
\usepackage{subcaption}

\newcommand{\approxprop}{\mathrel{\vcenter{
  \offinterlineskip\halign{\hfil$##$\cr
    \propto\cr\noalign{\kern2pt}\sim\cr\noalign{\kern-2pt}}}}}

\DeclareMathOperator{\sgn}{sgn}

\title[Eccentric debris disc morphologies I: face-on discs]{Eccentric debris disc morphologies I: exploring the origin of apocentre and pericentre glows in face-on debris discs}

\author[E. M. Lynch et al.]{
Elliot M. Lynch$^{1,2}$ \thanks{E-mail: elliot.lynch@ens-lyon.fr} and
Joshua B. Lovell$^{3}$
\\
$^{1}$Department of Applied Mathematics and Theoretical Physics, University of Cambridge, Centre for Mathematical Sciences, \\ Wilberforce Road, Cambridge CB3 0WA, UK\\
$^{2}$Univ Lyon, Univ Lyon1, Ens de Lyon, CNRS, Centre de Recherche Astrophysique de Lyon UMR5574, F-69230, Saint-Genis,-Laval, France \\
$^{3}$Institute of Astronomy, University of Cambridge, Madingley Road, Cambridge, CB3 0HA, UK
}

\date{Accepted XXX. Received YYY; in original form ZZZ}

\pubyear{2021}

\begin{document}
\label{firstpage}
\pagerange{\pageref{firstpage}--\pageref{lastpage}}
\maketitle

\begin{abstract}
The location of surface brightness maxima (e.g. apocentre and pericentre glow) in eccentric debris discs are often used to infer the underlying orbits of the dust and planetesimals that comprise the disc.
However, there is a misconception that eccentric discs have higher surface densities at apocentre and thus necessarily exhibit apocentre glow at long wavelengths. This arises from the expectation that the slower velocities at apocentre lead to a ``pile up'' of dust, which fails to account for the greater area over which dust is spread at apocentre.
Instead we show with theory and by modelling three different regimes that the morphology and surface brightness distributions of face-on debris discs are strongly dependent on their eccentricity profile (i.e. whether this is constant, rising or falling with distance).
We demonstrate that at shorter wavelengths the classical pericentre glow effect remains true, whereas at longer wavelengths discs can either demonstrate apocentre glow or pericentre glow.
We additionally show that at long wavelengths the same disc morphology can produce either apocentre glow or pericentre glow depending on the observational resolution. Finally, we show that the classical approach of interpreting eccentric debris discs using line densities is only valid under an extremely limited set of circumstances, which are unlikely to be met as debris disc observations become increasingly better resolved.
\end{abstract}

\begin{keywords}
planetary systems - circumstellar matter - submillimetre: planetary systems - celestial mechanics
\end{keywords}


\section{Introduction}
Debris discs are collisionally dominated belts of planetesimals that are observed around ${\sim}20\%$ of main sequence stars \citep{Wyatt08, Hughes18}. 
Their morphologies are diverse with many having now been observed with large scale asymmetries inferred as due to mechanisms such as recent planetesimal collisions \citep[e.g., in Beta Pictoris and HD10647, see ][]{Matra17, Lovell21C}, or from having an underlying eccentric distribution of planetesimals \citep[e.g., Fomalhaut, HR4796, HD202628, and HD38206 see][]{MacGregor13, Olofsson2019, Faramaz19, Booth21}. 
Such eccentric distributions can form through interactions between planetesimals and planets over ${\sim}$Myr-Gyr timescales \citep{Mustill09}; thus understanding the structure of asymmetric debris discs is important to correctly interpret the origin of planetesimal belt morphologies. 

Due to the structure of eccentric debris discs, an individual apse may become preferentially brighter than the other, an effect that has been referred to as pericentre glow \citep{Wyatt99} or apocentre glow \citep{Pan16} depending on which apse is enhanced.
At short wavelengths, dust emission is dominated by variations in the dust temperature, thus particles closer to the star in the pericentre direction become preferentially brighter.
At longer wavelengths however, variations in the surface density of dust become important. 
Consequently, it has been proposed that this leads to the preferential brightening of particles in the apocentre direction, which has arisen from the misconception that slower particle velocities at apocentre lead to dust pile ups, i.e., the `apocentre hang time' argument\footnote{This was originally introduced by \citet{Tremaine95} to clarify their asymmetric model of the M31 nuclear star cluster. In fact the original eccentric nuclear disc models \citep{Tremaine95,Statler99a,Statler99,Peiris03} required the eccentricity to \textit{decrease} outwards (bunching up orbits at apocentre) to produce the observed surface brightness maxima. However, a disc of stars with \textit{constant} eccentricity will have a constant surface brightness around the orbit, as pointed out by \citet{Statler99}.}.

Although previous studies which invoke pericentre glow to explain eccentric disc morphologies at short wavelengths remain valid for all discs we consider in this work \citep[e.g.,][]{Wyatt99}, in the long-wavelength limit the situation becomes more complicated, i.e., for cool debris discs observed at longer wavelengths. 
\citet{Pan16} modelled apocentre glow by considering \textit{linear} dust density distributions \citep[an issue discussed in][]{Marino19}, and is an extension of the apocentre hang time argument \citep[also invoked by][to explain the asymmetric debris disc of Fomalhaut]{Marsh05}. 
In this work we instead model the emission of debris discs with a \textit{surface} density distribution, which results in both qualitatively and quantitatively different predictions.

Throughout we use a number of different terms: `pericentre glow', `apocentre glow', `pericentre integrated flux enhancement', and `apocentre integrated flux enhancement'. The terms described as `glows' refer to the apse in which the peak \textit{surface brightness} appears, whereas the terms described as flux enhancements are measurements of which side of the disc's latus rectum contains more \textit{total flux} (i.e., integrated over an area of a disc).

In this paper we only consider face-on discs and neglect free eccentricity. We consider a disc with an eccentricity that is allowed to vary in such a way that there is no orbital intersection, as can occur in a disc with only forced eccentricity. Despite the lack of orbital crossings, the eccentricity we consider can be large, but is constrained to be less than unity. We adopt a simple prescription for the size distribution \citep[i.e. for $\alpha=-3.5$,][]{Dohnanyi69}, which does not vary in the disc. The neglected effects of disc inclination, free eccentricity and grain size dependant orbits will complicate considerably the results presented here, which we endeavour to explore in future work. Despite these rather restrictive assumptions we shall show that even the relatively simple discs we consider here can have complicated morphologies with observational consequences.

In Section \ref{sec:Geometry} we describe the geometry of an eccentric disc, in Section \ref{sec:Theory1} we present a model for a face on eccentric disc and derive expressions for the disc surface brightness. In Section \ref{sec:model} we describe the setup of the radiative transfer model which, in Section \ref{sec:discussion}, we use to verify earlier theoretical discussion over a range of wavelength regimes. We present our conclusions in \ref{sec:conclusions} and mathematical derivations are given in the appendices.

\section{Geometry of Eccentric Discs}
\label{sec:Geometry}
To describe an eccentric debris disc it is convenient to adopt an orbital coordinate system, which has a long history in celestial mechanics \citep[e.g.][]{Murray99}. As we are considering an extended disc we require the orbits of the disc particles to vary with their position in the disc. In a fluid disc the fluid orbits are a set of nested, nonintersecting Keplerian ellipses. In a debris disc there is no requirement that the particle orbits are non-intersecting, however this property will be kept due to its mathematical convenience. The continuous non-intersecting orbital coordinates systems have had some use in debris disc theory, notably \citet{Davydenkova18} who used one to study the secular forces associated with the disc self-gravity. 

There are many ways of constructing an orbital coordinate system (e.g. see \citet{Ogilvie01,Ogilvie14} for a different construction), in this work we shall follow \citet{Ogilvie19}.

Let $(x,y,z)$ and $(r,\phi,\tilde{z})$ be Cartesian and cylindrical polar coordinates, centred on the star, with $(x,y)$ and $(r,\phi)$ lying in the orbital plane, related by
\begin{equation}
x=r \cos \phi, \quad y = r \sin \phi, \quad z = \tilde{z} \quad .
\end{equation}

\begin{figure*}
\includegraphics[trim=100 245 100 280,clip,width=\linewidth]{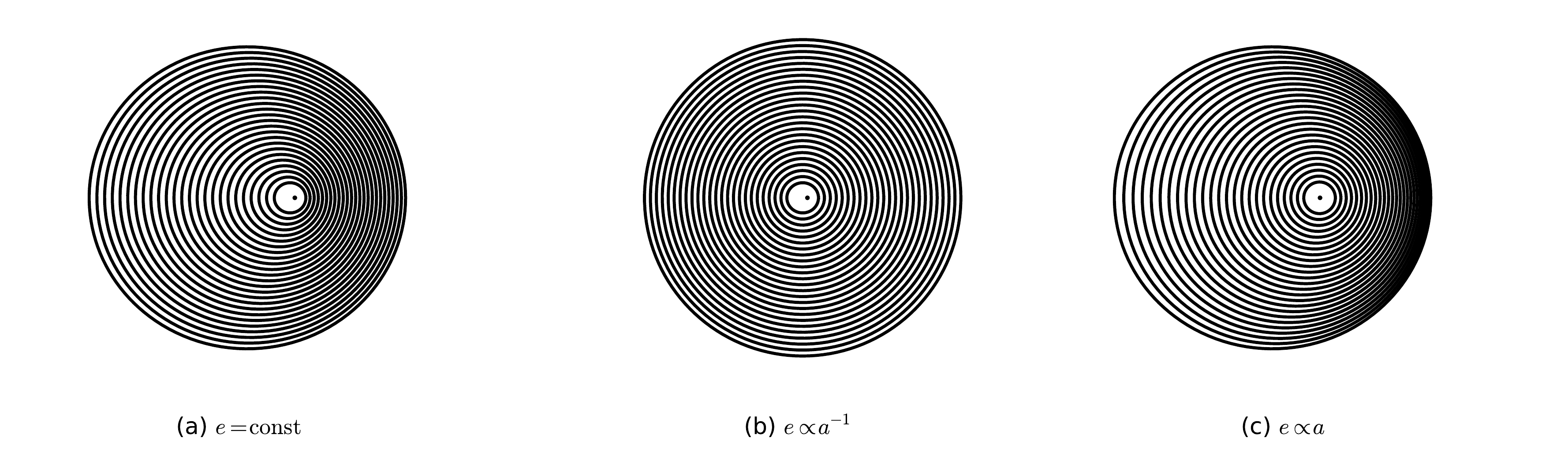}
\caption{Orbits in an (apse aligned) eccentric disc. Left: Constant eccentricity. Centre: $e \propto a^{-1}$. Right: $e \propto a$. The $e \propto a$ case shows a clear bunching of orbits at \textit{pericentre}, whereas a similar bunching of orbits at \textit{apocentre} occurs for $e \propto a^{-1}$, but not for the constant e case.}
\label{fig:orbits}
\end{figure*}

The polar equation for an elliptical Keplerian orbit of semimajor axis $a$, eccentricity $e$ and longitude of pericentre $\varpi$ is 
\begin{equation}
r = \frac{a (1 - e^2)}{1 + e \cos f} \quad,
\end{equation}
where $f = \phi - \varpi$ is the true anomaly. Equivalently, 
\begin{equation}
r = a (1 - e \cos E) \quad,
\end{equation}
where $E$ is the eccentric anomaly which satisfies: 
\begin{equation}
\cos f = \frac{\cos E - e}{1 - e \cos E}, \quad \sin f = \frac{\sqrt{1 - e^2} \sin E}{1 - e \cos E} \quad,
\end{equation}
as well as Kepler's Equation,
\begin{equation}
M= E - e \sin E \quad,
\end{equation}
where $M=n (t - \tau)$ is the mean anomaly, $n = (G M_1/a^3)^{1/2}$ is the mean motion, $M_1$ is the central mass and $\tau$ is the time of pericentre passage. We can describe the shape of the disc by considering $e$ and $\varpi$ to be functions of $a$. The derivative of these functions are written as $e_a = de/da$ and $\varpi_a=d \varpi/da$ (the latter corresponding to the disc twist). Figure \ref{fig:orbits} shows examples of eccentric discs constructed in this manner.

This eccentric disc model assumes that the dominant motion in the disc consists of elliptical Keplerian orbits, subject to relatively weak perturbations from secular forces. The secular perturbation will cause the orbital elements to slowly evolve and are responsible for determining how $e$ and $\varpi$ depend on $a$. Solving for the secular evolution of the disc is beyond the scope of this work, however the disc geometry we adopt is very flexible (excepting the limitation that $e(a)$ and $\varpi(a)$ be single valued in $a$) and allows for $e$ and $\varpi$ to be self-consistently determined through secular disc dynamics. In this work we shall make use of eccentricity profiles that were determined through secular interactions between test-particles and a companion \citep{Wyatt99,Wyatt05}. 

As per \cite{Ogilvie19} we define
\begin{equation}
q^2 = \frac{(a e_a)^2 + (1 - e^2) (a e \varpi_a)^2}{[1 - e (e + a e_a)]^2} \quad ,
\end{equation}
with the requirement that $|q| < 1$ to avoid orbital intersection. Similar criteria were derived by \citet{Statler01} and \citet{Ogilvie01}. The relative contribution of the eccentricity gradient and twist to $q$ is determined by an angle $\alpha$ with
\begin{equation}
\frac{a e_a}{1-e (e + a e_a)} = q \cos \alpha, \quad \frac{\sqrt{1 - e^2} a e \varpi_a}{1 - e (e + a e_a)} = q \sin \alpha \quad .
\end{equation}

The Jacobian determinant of the orbital coordinate system can be written in the form $J=J^{\circ} j (1 - e \cos E)$, where $J^{\circ} = a$ and 
\begin{equation}
j = \frac{1 - e (e + a e_a)}{\sqrt{1-e^2}} (1 - q \cos(E-\alpha)) \quad .
\end{equation}

This highlights the reason for maintaining the nonintersection criterion for our orbital coordinates. If there was an orbital intersection at some location in the disc (corresponding to $|q| \ge 1$) the Jacobian of the coordinate system will vanish at this location resulting in a coordinate singularity. This is due to the coordinates (in particular the semimajor axis $a$) being multivalued at this location. A disc without an orbital intersection can have a large forced eccentricity, but must have negligible free eccentricity (see, as an example, Figure \ref{fig:orbits}). Recently \citet{Kennedy20} suggested that the narrowness of Fomalhaut and HD202628 is indicative of precisely this scenario.
We note that in principle, one can include free eccentricity by introducing multiple (formally infinite) overlapping orbital coordinate systems and adding up the contributions to the surface density from particles in the separate coordinate systems.

\section{A Simplified Model of a Face on Disc}
\label{sec:Theory1}

In this section we shall consider a continuous 2D disc with no overlapping orbits, in which particles lie on a set of non-overlapping confocal Keplerian ellipses. For a particulate disc, eccentricity can be multivalued in semimajor axis, we also expect a distribution of inclinations and longitude of ascending nodes, causing a disc to have a finite thickness, so this model is an oversimplification of reality. However, for a nearly face on disc sculpted by a planet, this is a convenient toy model which highlights much of the salient features of a real disc. The planet will sculpt the disc in such a way that disc particles ought to lie close to the orbits considered. While we neglect it in this paper, the presence of a nonzero free eccentricity is an additional complication affecting debris disc morphology. A discussion of the influence of free eccentricity on the morphology of eccentric dust discs can be found in \citet{Kennedy20}.

For a nearly face on disc the 3D structure becomes less important as, even though the disc thickness will vary around the orbit, the same amount of mass is summed over when integrating along the line of sight to get the surface density/surface brightness. However the finite disc thickness will modify the average temperature around the orbit as material above the mid-plane will be slightly further from the star. As discs are expected to be thicker at apocentre this leads to a greater temperature contrast, between apocentre and pericentre, in 3D discs compared to 2D discs, which maybe important at shorter wavelengths. One can show, however, that at long-wavelengths the surface brightness of a disc with constant aspect ratio is proportional to that of a 2D disc, so the 3D structure is unimportant in this limit.

\subsection{Derivation of the model}

In the absence of any perturbing forces the disc material obeys mass conservation. As, in this simplified approximation, there are no overlapping orbits we can treat the dust as a fluid which obeys the continuity equation,
\begin{equation}
\dot{\rho} + \nabla \cdot \rho u = 0 \quad ,
\end{equation}
where an overdot indicates a partial time derivative. For a steady disc we require $\dot{\rho} = 0$, we define the surface density to be $\Sigma = \int \rho d z$ which obeys
\begin{equation}
\nabla \cdot (\Sigma u) = \frac{1}{J} \partial_{E} (J u^{E} \Sigma) = 0 \quad ,
\end{equation}
where $J$ is the Jacobian determinant of the orbital coordinate system and $u^{E} = n (1 - e \cos E)^{-1}$ is the velocity along the orbit. This can be straightforwardly integrated to obtain
\begin{equation}
\Sigma = \frac{M_a}{2 \pi a j} \quad,
\label{sigma eq}
\end{equation}
where $M_a$ is the mass per unit semimajor axis and is constant on each orbit. In the presence of eccentricity gradients or disc twist, $j$ will vary around the orbit due to the lateral compression of the test particle orbits. This lateral compression leads to a variable surface density.

For a dust disc consisting of a single grain size, with cross sectional area $\sigma_{\rm dust}$, that is a perfect absorber and emits as a black body, the surface brightness, in the optically thin regime,  can be found by multiplying the fractional cross sectional area of the dust (equal to $S_{\rm dust}/A = n H \sigma_{\rm dust}$, where $S_{\rm dust}/A$ is the total cross sectional area of the dust per unit area and, in this paragraph only, $n$ is the dust number density) by Planck's function \citep{Wyatt02},
\begin{equation}
I = B_{\lambda} (T) \sigma_{\rm dust} H n \propto B_{\lambda} (T) \Sigma \quad,
\label{surface brightness B sigma eq}
\end{equation}
where the Planck's function, $ B_{\lambda} (T)$, expressed in terms of the wavelengths is given by
\begin{equation}
B_{\lambda} = \frac{2 h c^2}{\lambda^5} \frac{1}{e^{\frac{h c}{\lambda k_B T}} - 1} \quad .
\end{equation}

The dust temperature is set by the black body equilibrium temperature given the stellar flux. For a star of radius $R_{*}$ and surface temperature $T_{*}$ the temperature in the disc, assuming spherical dust grains, is given by
\begin{equation}
T = \left(\frac{R_{*}}{2 R} \right)^{1/2} T_{*} \quad .
\end{equation}
where $R^2 = r^2 + z^2$ is the spherical radial distance from the origin. In this purely 2D model we neglect the vertical structure and set $R = r$. It is possible that the vertical temperature variation is important at shorter wavelengths. While we neglect this effect in our analytic models, our radiative transfer models (see $\S$\ref{sec:model}) account for this.

Normalising by a reference circular disc, where $\Sigma^{\circ} = \frac{M_a}{2 \pi a}$ is the surface density of the reference disc, we obtain a dimensionless surface brightness,
\begin{equation}
\mathcal{I} = \frac{ B_{\lambda} \left(T_{*} \sqrt{\frac{R_{*}}{2 r}}\right)}{j B_{\lambda} \left(T_{*} \sqrt{\frac{R_{*}}{2 a}}\right)} \quad,
\end{equation}
or
\begin{equation}
\mathcal{I} = j^{-1} \frac{e^{\frac{h c}{\lambda k_B T_{*}} \sqrt{\frac{2 a}{R_{*}}}} - 1}{e^{\frac{h c}{\lambda k_B T_* } \sqrt{\frac{2 a (1 - e \cos E)}{R_{*}}}} - 1} \quad .
\end{equation}
By introducing a characteristic wavelength $\lambda_{*}$, which we define as
\begin{equation}
\label{wav regime}
\lambda_{*} := \frac{hc}{k_{B} T_{*}} \left( \frac{2 a}{R_{*}} \right)^{1/2} \quad,
\end{equation}
the non-dimensional surface brightness becomes
\begin{equation}
\mathcal{I} = j^{-1} \frac{e^{\lambda_{*}/\lambda} - 1}{e^{\lambda_{*} (1 - e \cos E)^{1/2}/\lambda} - 1} \quad .
\label{surf bright eq}
\end{equation}
Thus we see that there are two contributions to the variation of the disc surface brightness around the orbit. One is the variation of the disc temperature in Planck's function as a result of the varying radial distance from the star, the second is the variation of the (dimensionless) Jacobian determinant $j$ (describing the variation of the disc surface density), that arises when there are lateral compressions in the dust orbits (i.e., from eccentricity gradients or disc twists).

To aid with analysis we define the surface brightness ratio $f$ to be the ratio of the surface brightness at apocentre to that at pericentre as
\begin{equation}
f := \frac{\mathcal{I} |_{E=\pi}}{\mathcal{I} |_{E=0}} \quad .
\end{equation} 
Similarly we define a total flux ratio $F$ to be the ratio of the total flux on the apocentre half of the disc to the total flux on the pericentre half of the disc as
\begin{equation}
F := \frac{\iint_{\mathcal{D}^a} I J \, d a d E}{\iint_{\mathcal{D}^p} I J \, d a d E } \quad ,
\end{equation}
where $\mathcal{D}^a$ is the subset of the disc with $\pi/2 < f < 3 \pi/2$ and $\mathcal{D}^p$ is the subset with $-\pi/2 < f < \pi/2$.

\subsection{Short wavelength limit, $\lambda \ll \lambda_{*}$, and the classic pericentre glow} \label{peri glow}

The surface brightness of a face on eccentric disc observed at short wavelength (or alternatively low disc temperatures) is well understood and results in the pericentre glow phenomena \citep{Wyatt99}. By considering the short wavelength limit (i.e., where $\lambda/\lambda_{*}~\ll~1$), Equation \ref{surf bright eq} becomes
\begin{equation}
\mathcal{I}_{\lambda \ll \lambda_{*}} \propto \frac{1}{1 - q \cos (E-\alpha)} \exp \left[\frac{\lambda_{*}}{\lambda} \left(1 - \sqrt{1 - e \cos E} \right) \right] \quad .
\end{equation}

Except where the disc particle orbits are very close to an orbital intersection ($q \approx 1$) the variation in the first term from the Jacobian will be much smaller than that from the exponential term which is due to the variation of the disc temperature around the orbit. The exponential term has a maxima at $E=0$ and we recover the classic pericentre glow effect.

We emphasise here that this is the expression for a razor thin disc (i.e., for a disc with zero scale height). There is an additional correction from the disc vertical structure owing to the greater temperature contrast between apocentre and pericentre in a 3D disc. Due to the sensitivity of Planck's function to temperature this correction could still be important even in thin discs. 

\subsection{Behaviour at Longer wavelengths} \label{long wavelength theory}

At longer wavelengths (or for warmer discs) the variation in the surface density around the orbit becomes important. In such a regime, the location of the emission maxima depends sensitively on the disc orbit geometry and the wavelength of observation. 

The simplest eccentric disc we can consider is an untwisted disc with constant eccentricity. For this disc $q=0$ which means $j$ and thus the surface density (by Equation \ref{sigma eq}) are independent of eccentric anomaly making them constant on each orbit. This can be understood as a direct consequence of Kepler's 2nd law which ensures that area is conserved by the orbital motion; as surface density is a reciprocal area, it too will be constant around the orbit. Consider now the surface brightness, our Equation \ref{surf bright eq} demonstrates that the only term to vary around the orbit is Planck's function (due to the temperature variation). This has a maxima at pericentre for all wavelengths. Therefore, a face on disc with constant eccentricity will always be brightest at pericentre, regardless of temperature or wavelength of observation. This contradicts the expectation of \citet{Pan16} (based on the apocentre hang time argument) which does not take into account the larger area over which the particles are spread at apocentre, and is thus an inappropriate model for a well resolved disc. This result does however agree with \citet{Marino19} where this effect was accounted for.

Next we consider the effect of eccentricity gradients in an untwisted disc. These cause the orbits to bunch up at apocentre (for $e_a < 0$) or at pericentre (for $e_a > 0$) causing an increase in the surface density (and consequently surface brightness) at these locations. The importance of the surface density variation to the surface brightness is controlled by the ratio $\lambda/\lambda_{*}$. When $\lambda/\lambda_{*} \ll 1$ surface density variations are much less important and, as discussed in Section \ref{peri glow}, produce classic pericentre glow discs. In the opposite limit ($\lambda/\lambda_{*} \gg 1$) the surface density has a greater effect on the surface brightness.

\begin{figure}
\includegraphics[width=\linewidth]{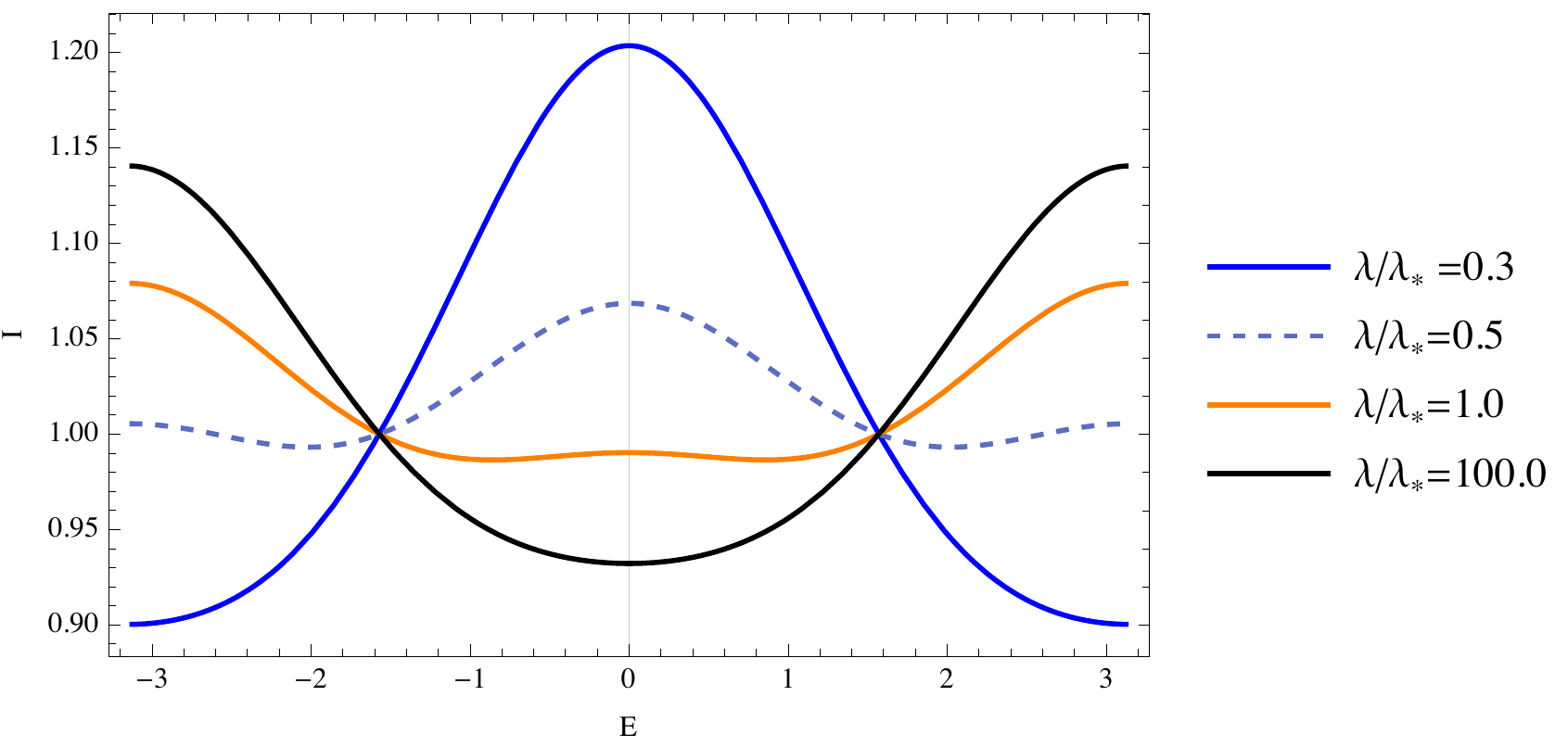}
\caption{Variation of the surface brightness around the orbit for $e{=}0.2$, $q=-0.2$ at different wavelengths, as might be expected for a disc with an internal perturber. At short wavelengths this has pericentre glow, however as the wavelength increases, the apocentre emission begins to dominate, with a double maxima at intermediate wavelengths and apocentre glow at long wavelengths.}
\label{fig:wavelength var}
\end{figure}

When $q < 0$, corresponding to the eccentricity decreasing outwards, the surface density is greatest at apocentre which competes with the higher temperatures at pericentre. For example, Figure \ref{fig:wavelength var} shows how, for an orbit with $e=0.2$, $q=-0.2$ (a situation which can occur in a disc with $e\propto a^{-1}$) the location of the brightest point in the disc changes from apocentre to pericentre as the wavelength is decreased. Interestingly at intermediate wavelengths we have two surface brightness maxima in the disc; one at apocentre associated with the enhanced disc \textit{surface density} and one at pericentre associated with the disc \textit{temperature}.

To explore the effect of density enhancement further we considered the limit $\lambda/\lambda_{*} \gg 1$ where the effects of the surface density are most apparent. The details of this limit are given in Appendix \ref{long wav limit}. Figure \ref{fig:e var} shows how the surface brightness changes around the orbit for different eccentricities in this limit when $q=-0.1$. This shows how the location of the surface brightness maxima depends on the orbital geometry when $\lambda/\lambda_{*} \gg 1$. Additionally it shows that in the long wavelength limit there is a band of eccentricities which produce the double maxima (as also seen at intermediate $\lambda/\lambda_{*}$ in Figure \ref{fig:wavelength var}).

In the long wavelength limit the ratio of apocentre surface brightness to pericentre surface brightness $f$ is given by
\begin{equation}
\lim_{\lambda/\lambda_{*} \rightarrow \infty} f = \frac{1-q}{1+q} \sqrt{\frac{1-e}{1+e}} \quad .
\label{i ratio}
\end{equation}
For an untwisted disc with constant eccentricity $q=0$ and $f < 1$. Closely related expressions were derived in \citet{Tremaine95}, \citet{Statler99} and \citet{Peiris03}; although, since the discs in these studies were composed of stars (i.e., not dust), the contribution from the variation of Planck's function around the orbit was not included.

\begin{figure}
\includegraphics[width=\linewidth]{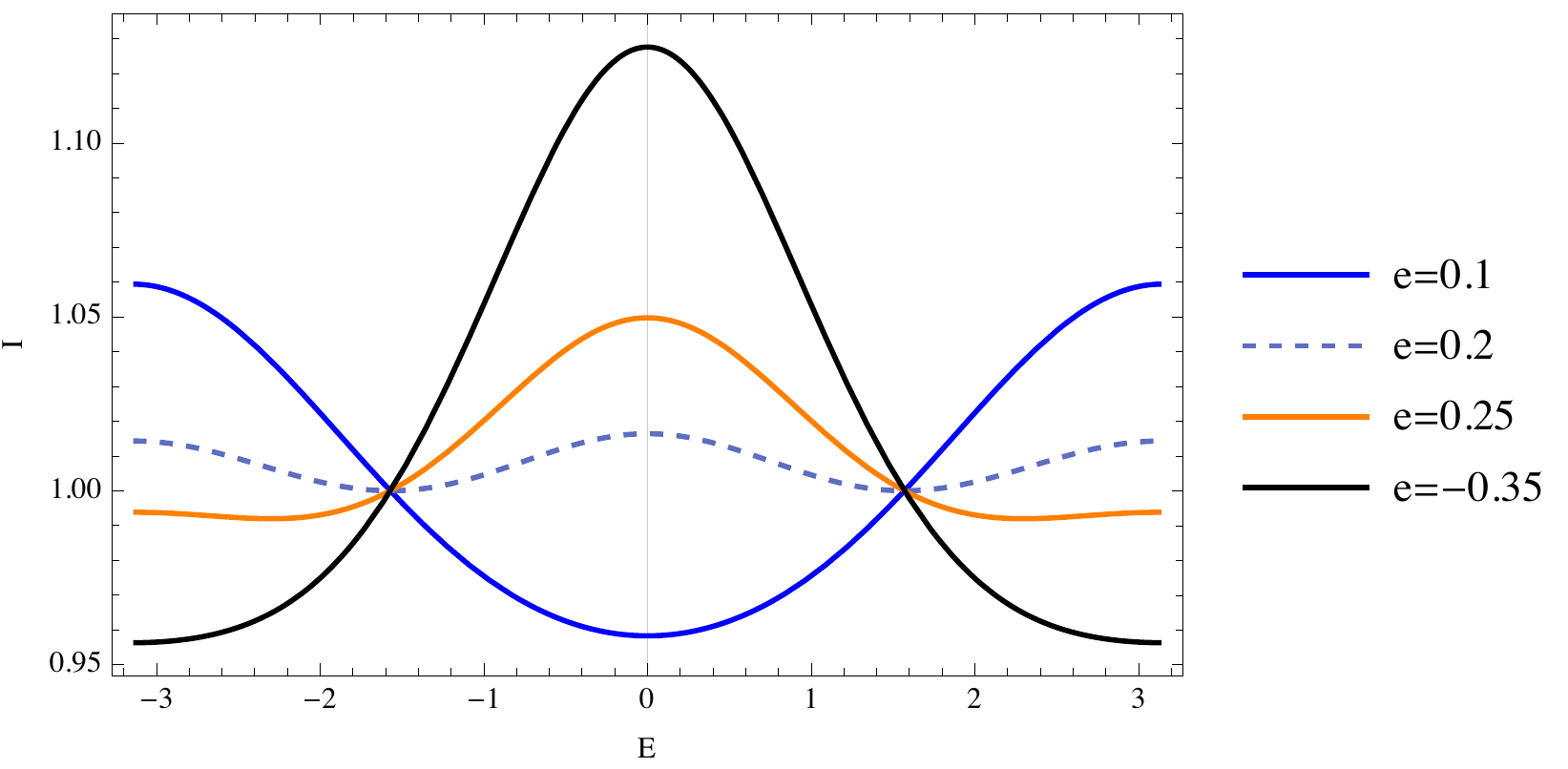}
\caption{Variation of surface brightness around the orbit for $q=-0.1$ for various eccentricities in the long wavelength limit.}
\label{fig:e var}
\end{figure}

Concretely if we have an eccentricity distribution $e \propto a^{-1}$ (i.e., as expected for the forced eccentricity from a perturber \textit{internal} to a debris disc) the ratio of $\mathcal{I}$ at apocentre to that at pericentre is
\begin{equation}
\lim_{\lambda/\lambda_{*} \rightarrow \infty} f^{\rm fall} = \sqrt{\frac{1+e}{1 - e}} \quad,
\label{inner perturber i ratio}
\end{equation}
which would result in apocentre glow. 
Interestingly this is the same apocentre to pericentre surface brightness ratio that would be predicted for a constant $e$ ring using the classical line density method. This suggests that where the classical line density model with constant $e$ has been shown to be consistent with observations of debris discs \citep[e.g.,][]{MacGregor17}, this may in fact be a misinterpretation of the effect of a negative eccentricity gradient, consistent with an internal perturber. We discuss this further in $\S$\ref{sec:resolution}, alongside the effect of resolution.

\begin{figure}
\includegraphics[width=\linewidth]{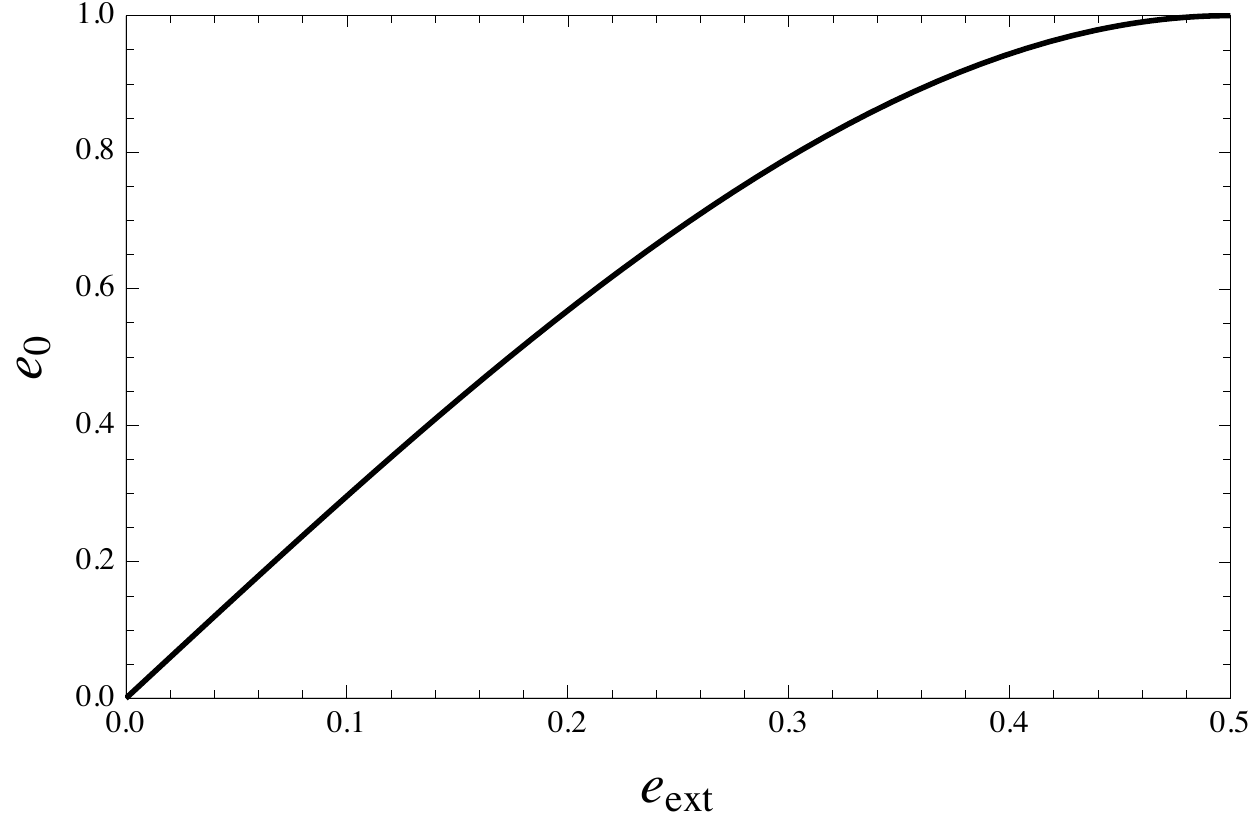}
\caption{Eccentricity obtained in the long wavelength limit by fitting a disc with $e \propto a$ (i.e., consistent with that resulting from an external perturber) with a constant eccentricity ring. $e_{\rm ext}$ is the actual eccentricity of the disc, while $e_0$ is the eccentricity that would be inferred if this was fitted with a constant eccentricity ring. Typically the inferred eccentricity is about twice the real eccentricity. At shorter wavelengths this discrepancy is reduced.}
\label{e overpredict}
\end{figure}

Considering now the case of a disc where the eccentricity \textit{increases} outwards (i.e., $q>0$, as expected for the forced eccentricity from a perturber \textit{external} to a debris disc). In this instance both the density and the temperature contribute to the surface brightness enhancement at pericentre, resulting in this becoming brighter than that in a disc with a constant eccentricity profile. An important consequence of the enhancement of pericentre glow is that fitting the debris disc with a constant eccentricity ring will result in an over-prediction of the disc eccentricity. Figure \ref{e overpredict} shows how fitting a disc with an external perturber and $e \propto a$ with a constant eccentricity ring can result in a factor of 2 over-prediction in the eccentricity. We can see the effects of this surface brightness enhancement by comparing the surface brightness ratio of this disc $f^{\rm rise}$ to that of a disc with constant eccentricity $f^{\rm const}$, that is
\begin{equation}
\lim_{\lambda/\lambda_{*} \rightarrow \infty} \frac{f^{\rm rise}}{f^{\rm const}} = \frac{1 - 2 e}{1 + 2 e} \frac{1+e}{1-e} \le 1 \quad .
\label{peri bright enhacement rise}
\end{equation}
This quantifies the enhancement of the pericentre glow by the eccentricity gradient for a given $e$, and shows that the smaller this ratio becomes, the larger the enhancement of the pericentre glow relative to the constant eccentricity disc.

We summarise our key findings of the behaviour of eccentric discs thus far in Table \ref{tab:eDefs}, which we shall explore further in our radiative transfer models (see $\S$\ref{sec:model}).
 
\begin{table}
\centering
\caption{Long wavelength behaviour for the three scenarios considered in this work. Note that `Int.' refers to internal, `Ext.' to external, `Peri.' to pericentre, `Apo.' to apocentre, and `Max.' to maxima.}
\begin{tabular}{ l | c c c}
\hline
\hline
 & Single Belt & Ext. Perturber & Int. Perturber \\
\hline
eccentricity & $e = \textrm{const}$ & $e = e_0 \left( \frac{a}{a_0} \right)$ & $e = e_0 \left( \frac{a_0}{a} \right)$  \\ 
q & $q = 0$ & $q = \frac{e}{1 - 2 e^2}$ & $q = -e$ \\
N$_{\rm{extrema}}$ & 2 & 2 & $\begin{cases} 2 \quad e {<} \frac{1}{3} \\ 4 \quad e {>} \frac{1}{3}  \end{cases}$ \\
Max. ($\lambda {\gg} \lambda_{*}$) & Peri. & Strong Peri. & $\begin{cases} \textrm{Apo.} \quad e {<} \frac{1}{3} \\ \textrm{Both} \quad e {>} \frac{1}{3}  \end{cases}$ \\
$\lim_{\lambda/\lambda_{*} \rightarrow \infty} f$ & $\sqrt{\frac{1-e}{1+e}} {\le} 1$ & $\frac{1 - 2 e}{1 + 2 e} \sqrt{\frac{1+e}{1-e}} {\le} 1$ & $\sqrt{\frac{1+e}{1 - e}} {\ge} 1$ \\
\hline
\end{tabular}
\label{tab:eDefs}
\end{table}

\subsection{Effects of finite resolution}
\label{sec:Theory2}
Thus far the models we have been considering have had infinite resolution, reflecting the real surface brightness distribution of the disc. In practice all observations are carried out with a beam of finite size, which will affect the disc morphology observed. In what follows we shall only consider symmetric beams (however, the generalisation to asymmetric beams is straightforward and is provided in Appendix \ref{beam convolution explicit}). For a symmetric Gaussian beam,
\begin{equation}
B = \frac{1}{2 \pi b^2} \exp \left( -\frac{x^2 + y^2}{2 b^2} \right),
\end{equation}
then the observed surface brightness is obtained by convolving the disc surface brightness with the beam: 
\begin{equation}
I_{\rm obs} = I * B ,
\end{equation}
where $*$ denotes a 2D convolution.
Now, we write the disc mass per unit semimajor axis as
\begin{equation}
M_a = \frac{\sqrt{2 \pi}}{w} a \sigma_0(a) \exp \left(-\frac{(a - a_0)^2}{2 w^2} \right) ,
\end{equation}
where $w$ is a typical disc length scale (e.g., the disc width) and $\sigma_0(a)$ is arbitrary (as $\sigma_0$ is arbitrary we can always do this). For a relatively narrow beam (i.e., where $b \ll a_0$ is small enough such that $T(a,E)$, $\sigma_0 (a)$, $e (a)$ and $\varpi (a)$ are all approximately constant within a beam) the only variation of the disc surface brightness within a beam will be due to the $\exp\left(-\frac{(a - a_0)^2}{2 w^2}\right)$ term in $M_a$, which becomes important when $w \lesssim b$. We only consider the long-wavelength limit here (as otherwise the variation of the disc surface brightness due to temperature becomes too strong to treat the effects of temperature as being constant within a beam). Under these assumptions we can approximate the convolution using Laplace's method (for which we provide a detailed derivation in Appendix \ref{beam convolution explicit}) to obtain
\begin{align}
\begin{split}
I_{\rm obs} &\approxprop \frac{1}{\sqrt{2 \pi} w j}  T(a,E) \sigma_0(a) \left(1 + \frac{b^2 (1 + e \cos E)}{w^2 j^2 (1 - e \cos E)} \right)^{-1/2} \\
&\times \exp \left[ - \frac{(a - a_0)^2}{2 w^2} \left(1 + \frac{b^2 (1 + e \cos E)}{w^2  j^2 (1 - e \cos E)} \right)^{-1}\right] .
\end{split}
\label{beam convolved I}
\end{align}
In the limit $b \ll w$ we recover the infinite resolution result,
\begin{equation}
I_{\rm obs} \approxprop \frac{1}{\sqrt{2 \pi} w j} T(a,E) \sigma_0 (a) \exp \left( - \frac{(a - a_0)^2}{2 w^2} \right) ,
\end{equation}
whereas in the opposite limit (i.e., $w \ll b$, where the disc width is much narrower than the beam but the the beam is much smaller than the azimuthal variation in the disc) we instead obtain 
\begin{equation}
\begin{split}
I_{\rm obs} \approxprop \frac{T(a_0,E) \sigma_0 (a_0)}{\sqrt{2 \pi} b } \sqrt{\frac{1 - e_0 \cos E}{1 + e_0 \cos E}} \\
      \times \exp \left( - \frac{j^2 (1 - e_0 \cos E)}{2 b^2 (1 + e_0 \cos E)} (a - a_0)^2 \right) ,
\label{unresolved surface brightness}
\end{split}
\end{equation}
where we have set $a=a_0$ except where it appears in the exponent.

At the ring peak $a=a_0$ we find $I_{\rm obs} \propto T \sqrt{\frac{1 - e \cos E}{1 + e \cos E}} \propto T/v$, which agrees with the prediction of \citet{Pan16}, however we note that the physical picture is somewhat different to the explanation provided there. The increased flux at apocentre is not caused by the surface density being higher at apocentre but is instead due to the ring being wider at apocentre. When the beam is significantly wider than the ring width this leads to more mass under the beam at apocentre than at pericentre.

Interestingly, while the ring peak at $a=a_0$ has $I_{\rm obs} \propto T/v$ (and is independent of the eccentricity gradient) the shape of the ring away from $a=a_0$ is sensitive to the eccentricity gradient. This means that it may be possible to extract the eccentricity gradient in an unresolved eccentric ring (e.g., by analysing the residual image emission after subtracting off the surface brightness predicted for a constant $e$ ring) particularly if the eccentricity of the ring can be independently measured from the stellar offset.

\subsection{Disc Twist}
\label{sec:Theorytwists}
The theory we have developed allows for the disc to be twisted, where the longitude of pericentre varies with semimajor axis. Disc twist ($\varpi_a \ne 0$) will shift the location of the maxima in the surface density distribution away from the line of apse. This can mean the surface brightness maxima is no longer located on the line of apse, a situation that has received little attention in the debris disc literature despite this having implications for the morphology of debris discs. For example, in the long wavelength limit for a highly twisted disc, the surface brightness maxima will be located at $E = \sgn(\varpi_a) \pi/2$ and can thus be at right-angles to the line of apse.

In fluid discs, disc twist is a natural outcome of dissipation within the disc, along with the transmission of eccentricity by eccentric waves. In a debris disc the main collective effects that could be present is collisional viscosity and the disc self gravity \citep[and these effects can produce twisted eccentric waves in planetary rings, see][]{Borderies83,Shu84,Borderies85,Borderies86}. 
A further example of the importance of debris disc twists was recently shown in the simulations of \citet{Sende19}, where the presence of a planet internal to a debris disc can give rise to such twists across a range of grain sizes. Additionally, disc twists can vary due to radiation pressure causing dust grains of different sizes to have different semimajor axes, resulting in differing precession rates from secular forces.

\subsection{Summary}
\label{theoretsummary}

In this section we have described a simplified model for a face on, 2D, debris disc in the absence of significant free eccentricity. This is a minimal model capable of accurately describing eccentric debris disc morphology. The surface brightness is given by Equation \ref{surface brightness B sigma eq}, which is a common approximation for studying optically thin debris discs. For the mass distribution we derived the \textit{surface density} for an eccentric disc (Equation \ref{sigma eq}) rather than the more commonly used \textit{line density}. 

While not new, Equation \ref{sigma eq} seems to have been overlooked by the debris disc community. We have explored the disc morphology resulting from this mass distribution and have shown that it is fundamentally different from that expected from a line density, which does not account for the changing disc area around an orbit. Finally we have derived an expression for a beam convolved surface brightness (Equation \ref{beam convolved I}) and have used it to show that the classical $I_{\rm obs} \propto T/v$ prediction from using a line density is only recovered when there is a separation of scales between the disc size $r$, beam size $b$ and ring width $w$ such that $r \gg b \gg w$.

In the next section we shall setup radiative transfer models using the dust densities obtained from Equation \ref{sigma eq} in order to test the predictions made here.

\begin{figure*}
    \includegraphics[width=2.0\columnwidth]{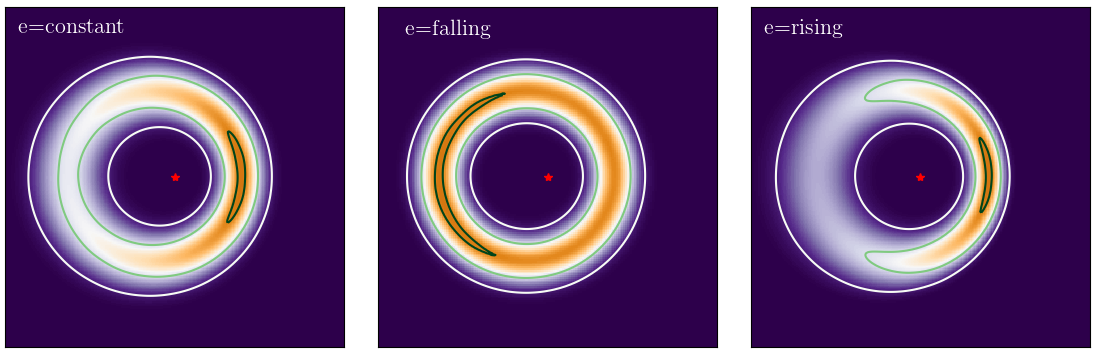}
    \caption{Images for the three models considered in this work (here observed with $\lambda{=}1\,$mm, for the case that $e_a(r_0)=0.25$. Visible in all are the location of the star (red stellar marker), and contour lines showing 10\%, 50\% and 95\% of the respective model peak surface brightnesses. These can be seen to be at a maximum on the pericentre side in the e=constant and e=rising models, and on the apocentre side in the e=falling model. The physical scale over which the peak emission dominates the apses varies for the different models, which is discussed further in $\S$\ref{sec:discussion}.}
    \label{fig:exp4}
\end{figure*}


\section{Analytical Disc Modelling}
\label{sec:model}
Following our theoretical derivations of debris disc morphologies in $\S$\ref{sec:Theory1}, we continue here by verifying these with known radiative transfer code. 
We developed three non-dynamical, Keplerian models to be investigated with the $\rm{RADMC-3D}$ \citep{Dullemond12} package, and used these as the basis of our investigation. Each of these models was parametrised as having a surface density, defined by Equation \ref{sigma eq}. 

Whilst our models investigate three different relationships between the emission of the disc and the gradient of the eccentricity, these all share a common set of parameters, many of which are fixed throughout our modelling investigation. 
Together, these define the thermal emission from a single, optically thin disc (for a given observation wavelength ($\lambda_{\rm{obs}}$) and observation resolution) with a Gaussian-distributed radial profile with distance ($d$, fixed at 10\,pc), peak emission semi-major axis ($a_0$), width ($w$, related to the FWHM of the disc via $\rm{FWHM}{=}2\sqrt{2\ln{2}}\,w$), vertical aspect ratio ($h$, fixed at 3\%), an eccentricity ($e(a)$, as defined in Table~\ref{tab:eDefs}), argument of pericentre ($\omega$, fixed at $0^\circ$, parallel with the disc major axis), inclination ($i$, fixed at $0^\circ$, i.e., face-on), position angle ($\rm{PA}$, from North, anti-clockwise, fixed at $-90^\circ$, i.e., parallel with the argument of pericentre), total dust mass ($M_{\rm{dust}}$, fixed at $0.05\,M_\oplus$), minimum and maximum grain sizes ($D_{\rm{min}}$ and $D_{\rm{max}}$, respectively set by the blowout size of a ${\sim}$solar-type star, and by neglecting emission from larger grains than ALMA wavelengths at values of $0.9\,\mu$m and 1.0\,cm), dust grain density ($\rho$, fixed at $2.7\,\rm{g\,cm}^{-3}$), and grain size power-law distribution \citep[$\alpha$, fixed at -3.5, as per][]{Dohnanyi69}. 
The dust temperature in our models is determined by the stellar temperature ($T_\star$) and stellar radius ($R_\star$) which define a template \citep{Kurucz79} stellar spectra (in all instances in this work these were fixed as $7000\,$K and $1.2\,R_\odot$ respectively), and the stellar mass ($M_\star$) was fixed as $1.1\,M_\odot$. 
In all images however we re-scale the flux of the star to $F_{\rm{star}}=100\,\mu\rm{Jy}$ (consistent with the stellar flux of stars with these parameters at $\lambda {\sim} 1\,$ mm), and fix the origin of the coordinate system at the star's location (we note for the purposes of our investigation, the stellar flux is negligible in comparison to the disc flux, and thus has no influence on the modelled debris disc morphologies).
Our models have a vertical Gaussian density distribution, defined by the vertical aspect ratio, $h=H/r$, where $H$ is the absolute vertical height of dust at a radius $r$ in the disc. 
This same parameterisation has been applied previously to model sub-mm ALMA observations \citep[see][]{Marino16, Marino18, Lovell21C}, and is consistent with the expected physical distribution of dust above and below the disc mid-plane.
It is possible to show that this vertical structure is a natural consequence of the inclination distribution in the disc \citep[see, for example, ][]{Matra19}.

In conjunction, these parameters ensure that all discs remain optically thin, and thus comparable with commonly observed debris discs (e.g., with radii and widths in the tens of au, and with disc fluxes in the tens of mJy at nearby ${\sim}10$\,pc distances). We note however for denser discs (e.g., optically thick protoplanetary discs) the situation diverges from our theory.
In Fig.~\ref{fig:exp4} we demonstrate three images produced by the three template models (for $e(a_0)=0.25$) in which the outcomes described in Table~\ref{tab:eDefs} are apparent. 
Our models are projected on to a 2-D grid in $(r,\phi)$ space, where $r=0.0''$ corresponds to the origin (i.e., the location of the star), and both $\omega=0^{\circ}$ and $\phi=0^{\circ}$ point along the major axis of the disc towards the pericentre. 


\section{Disc Modelling Results}
\label{sec:discussion}
In this section we outline the results of the three models introduced in $\S$\ref{sec:model}, in relation to the asymmetric surface brightness and total flux enhancements these can produce, as described functionally in Table~\ref{tab:eDefs}, and with the parameters outlined in $\S$\ref{sec:model}. 
We define the baseline set of scenario parameters that all our models use (unless otherwise stated) as $a_0=70$\,au, $w=30$\,au and a resolution $=0.02''$ (i.e., 0.2\,au at the 10\,pc distance chosen). 
Based on Equation~\ref{wav regime}, such disc parameters ensure that at an observational wavelength of 10s of $\mu$m, our model discs are in the short wavelength limit, whereas at an observational wavelength of 1\,mm, our model discs are in the long wavelength limit.
In this section we introduce the measurements $f^{\rm{scen}}$ and $F^{\rm{scen}}$, which respectively refer to the apocentre-to-pericentre surface brightness ratio of the peak emission, and to the apocentre-to-pericentre integrated flux ratio between all modelled disc emission on either side of the line parallel to the latus rectum and through the star, for the scenarios where the eccentricity is either constant (scen=con), falling (scen=fall) or rising (scen=rise).

\subsection{Disc Brightness Enhancements}
\label{sec:discBrightness}
This section primarily considers the parameter $f^{\rm{scen}}$, which seeks to quantify the `glow' of discs, i.e., the ratio of the peak \textit{surface brightnesses} in the two apses for our modelled scenarios (i.e., the three models with different eccentricity gradients). 
We show in this section that at long wavelengths (such as those sub-mm/mm accessible with ALMA) debris discs can show either apocentre glow, or pericentre glow, depending on the morphology of their dust distributions.

We note here that disc twists (discussed in $\S$\ref{sec:Theorytwists}) can introduce dust density (and therefore surface brightness) enhancements away from apse in arbitrary locations in the disc.
Given this, we do not further consider the effect of disc twists in our modelling approach, but note that disc twists may be the origin of disc asymmetries, and can (if sufficiently dense) shift the location of disc peak surface brightnesses in \textit{any} given azimuthal angle around the disc. 

\begin{figure}
    \includegraphics[width=\linewidth]{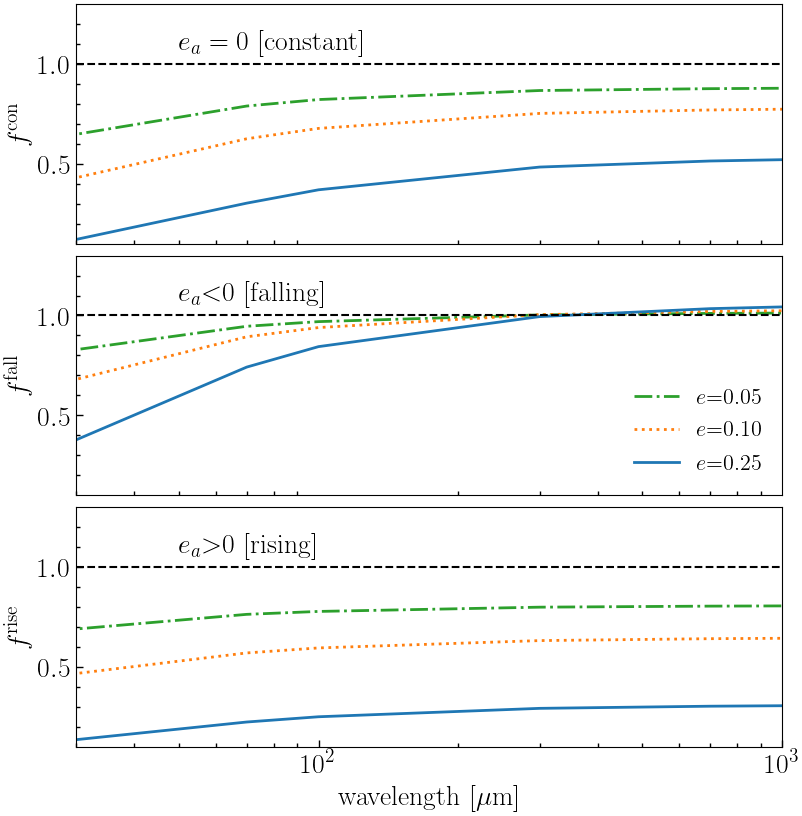}
    \caption{Plots to demonstrate the apocentre-pericentre surface brightness enhancements for discs with $e(a_0){=}0.05$, $0.10$ and $0.25$, at a resolution of 0.02''. Top: constant eccentricity discs, middle: falling eccentricity discs, bottom: rising eccentricity discs.}
    \label{fig:BrightRatios}
\end{figure}

\subsubsection{How and where can apocentre glows form?}
We first consider the scenarios that give rise to enhanced apocentre-to-pericentre surface brightness ratios, first with reference to Fig.~\ref{fig:exp4}, which shows three model scenarios with contours demonstrating the location of peak surface brightness, and the location of the star, first imaged at 0.02'' resolution at a wavelength of 1\,mm.
In all three scenarios the star is at a fixed location, and it can be seen that in only one scenario the apocentre brighter the pericentre - i.e., in the case where the eccentricity is falling as a function of the semimajor axis (whereas in both e=const and e=falling scenarios, the pericentre is brighter than their respective apocentre).

We explored how this glow varies with eccentricity (by choosing values of either 5\%, 10\% or 25\%, referenced at disc semi-major axis) and wavelength (ranging from the mid-IR to the sub-mm, $30\,\mu$m to 1\,mm) and show in the middle panel of Fig.~\ref{fig:BrightRatios} our results. 
It can be seen in all cases for the negative eccentricity gradient scenario, that enhanced apocentre surface brightness is achieved beyond 300-400\,$\mu$m (for all eccentricities) and that this continues to rise with increasing wavelength and eccentricity.
Although this is observed at the few percent level, this confirms the long-wavelength theory behaviour, as derived in $\S$\ref{long wavelength theory} for a nominal set of debris disc parameters, and is measurable within the errors of modern astronomical instrumentation (such as ALMA in the sub-mm). 
Since falling eccentricities with semi-major axis arises in the first-order expansion in eccentricity for the eccentricity of planetesimal belts with interior eccentric planets, this has important implications for interpreting planetary systems. For example, if observations of debris discs show significant apocentre glow, this could be used to infer the presence of \textit{internally} perturbing massive bodies - e.g., a massive planet inside a debris disc (making it possible to rule out the eccentricity of the belt being constant with semi-major axis, though we note that determining the precise parameters of a given perturber based on this surface brightness ratio requires care). On the other hand, apocentre glows can also be introduced as a result of the observational resolution, which we discuss further in $\S$\ref{sec:resolution}. 

\begin{figure*}
    \includegraphics[width=2.0\columnwidth]{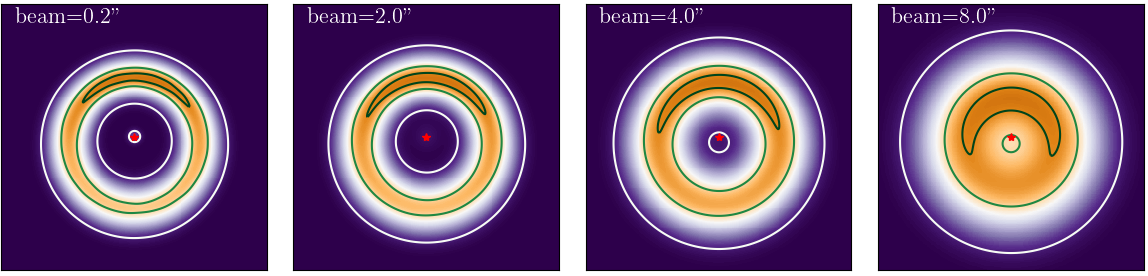}
    \caption{Images (shown at $\lambda=1\,$mm) to show for constant $e{=}0.1$ discs how the resolution of the imaging affects the location of the peak surface brightness, and disc centre, showing from left to right resolutions of 0.2'', 2.0'', 4.0'' and 8.0''. Image contours are shown for 10\%, 75\% and 95\% of the peak surface brightness.}
    \label{fig:expRes}
\end{figure*}

\subsubsection{How and where can pericentre glows form?} \label{sec:peri glow radmc}
We next explore how the effect of pericentre glows can form, by first referring to Fig.~\ref{fig:exp4} once more. 
This shows that enhanced surface brightnesses are located in the pericentre direction in the cases of the eccentricity gradient being either flat, or positive.
To add to this, we show in Fig.~\ref{fig:BrightRatios} in the upper and lower panels (for the same reference eccentricities and wavelengths as considered for the falling eccentricity scenario) how these vary for 0.02'' resolution images.
Most interesting in both of these plots, is that at none of the explored wavelengths or eccentricities does the surface brightness at apocentre ever exceed that at pericentre, even in the long-wavelength regime.
This also shows that this pericentre surface brightness is increased for the highest eccentricities, at shortest wavelengths \citep[consistent with the classical pericentre glow theory discussed by][]{Wyatt99}, and for increasing eccentricity gradients. 
This conflicts with the result of \citet{Pan16}, however we again note the key difference between our studies being that here we have modelled disc \textit{surface} densities, whereas in \citet{Pan16} apocentre glow predictions were made based on ring \textit{line} densities (which we showed in $\S$\ref{long wavelength theory} will result in quantitatively different predictions).
Since a linearly rising eccentricity with semi-major axis corresponds to the first-order expansion in $e$ for the eccentricity for secularly perturbing external bodies this has important implications for interpreting planetary systems. For example, if a significant pericentre glow was measured in sub-mm observations of a debris disc \citep[in the absence of other disc sub-structures, e.g., clumps inside the belt such as in the case of HD10647, see][]{Lovell21C}, this could be consistent with (or due directly to) perturbations from an external eccentric body (such as a massive planet orbiting outside the disc).
To add to this final point, whilst pericentre glows can be set up in either e=constant or e=rising discs, by comparing the upper and lower plots of Fig.~\ref{fig:BrightRatios}, it can be seen that in systems where the eccentricity rises with radius, the pericentre glows are much higher than those for constant eccentricity discs in the long wavelength regime, by as much as 10s of percent. Such a difference may therefore allow us to better determine the distribution of eccentricities in the disc, and thus whether or not this is indeed consistent with an external perturber scenario.

\subsubsection{Effect of resolution}
\label{sec:resolution}
For a single constant eccentricity model (setup identically as in $\S$\ref{sec:peri glow radmc}) with $e_0=0.10$, we considered how the image resolution can affect the surface brightness distribution. We show in Fig.~\ref{fig:expRes} this model convolved with four different beams (at resolutions of 0.2'', 2.0'', 4.0'' and 8.0''), and note have all been rotated $90^{\circ}$ anti-clockwise such that the $\rm{PA}$ is pointing upwards. 
There are two important effects of note. 
Firstly that simply observing a constant $e$ disc at a low enough resolution is not sufficient to produce sub-mm apocentre glow, i.e. the underlying disc distribution is important. Secondly, in the highly unresolved situations (i.e., the 4.0'' (40\,au) and 8.0'' (80\,au) images) the stellar location diverges from the central minima. 
We thus note here that this may be important for low-resolution image analysis when the position of the star may only be known with limited precision, since the distance between the central minima and apocentre peak is less than the central minima to pericentre peak, this could result in confusion between the direction of pericentre, despite the absolute position of the star not changing.

\begin{figure}
    \includegraphics[width=\linewidth]{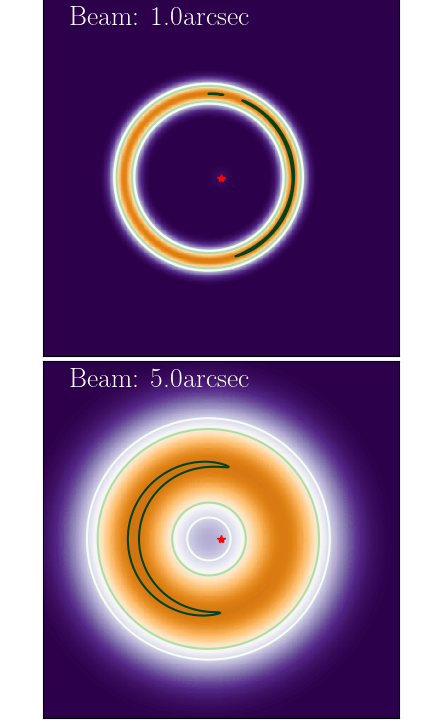}
    \caption{Images (shown at $\lambda=1\,$mm) show that either apocentre or pericentre glows can be introduced in the same discs based on resolution, shown here for the case of a narrow ($w=0.1r_0$) constant eccentricity (here $e=0.15$) 50\,au disc. Top: pericentre glow introduced with a 1.0'' beam, and bottom: apocentre glow introduced by 5.0'' beam. In both, contours show the 50\%, 65\% and 99\% emission levels (scaled to the image peak).}
    \label{fig:apoPeriResln}
\end{figure}

As discussed in Section \ref{sec:Theory2} there is, however, a low resolution, long-wavelength, limit where the classical expectation of apocentre glow holds, namely when $w \ll b \ll a_0$. We show in Figure \ref{fig:apoPeriResln} this effect by considering a narrow eccentric disc (i.e., with a width of just 10\% of its peak emission radius, and $e=0.15$) where the same disc is imaged with 1'' (10\,au) and 5'' (50\,au) beams. This shows that purely by reducing the resolution, the brighter apse can flip from the pericentre direction to the apocentre direction at 1\,mm wavelengths. In this scenario, pericentre glow or apocentre glow can be introduced solely as a result of image resolution. 

Table \ref{tab:discRegimes} lists a number of discs that have been interpreted as being eccentric, and compares the disc sizes and widths against the beam size of observations with current instruments.
We see that for the ``narrow'' eccentric discs (for which we use the examples of Fomalhaut, HR\,4796, and HD\,202628) current observations have $b \sim w$; this falls outside the regime discussed above, and so an interpretation of the disc eccentricity based on equations derived from line densities is inappropriate, and may only be so when the beam size meets the above condition.
``Wide'' discs (of which HD\,38206 is just one example) in which disc radii and widths are comparable (i.e., $r \sim w$) always fall outside the regime discussed above; interpretations of their eccentricities based on equations derived from line densities are never appropriate, irrespective of the beam size.

Experimentation with varying the beam size for a face on disc with the inferred parameters of HR\,4796 shows that when it is poorly resolved (beam size $\sim$ double that of the ALMA observations presented in \citet{Kennedy18}) the observations sit in the regime where the classical line density method is valid.
The same is likely true of the other ``narrow'' discs in Table~\ref{tab:discRegimes}.
We thereby conclude that the classic line density method of interpreting disc eccentricities is only valid for certain types of discs (e.g., ``narrow''), and then only when observations meet the condition $w \ll b \ll a_0$.
As we increasingly improve the resolution of optically thin debris disc imaging at long wavelengths (with instruments such as ALMA), observations that span multiple beams across the widths of discs will be increasingly affected by dust density variations due to eccentricity gradients and therefore cannot be interpreted using line density model predictions.

As discussed in $\S$\ref{long wavelength theory} the surface brightness ratio for a disc with an internal perturber is identical to that predicted using the classical line density for the same eccentricity at the ring peak.
In fact using Equation~\ref{beam convolved I} we can show that a narrow ring with $e \propto a^{-1}$ has an apocentre to pericentre flux ratio $f = \sqrt{\frac{1 + e}{1 - e}}$ irrespective of $b/w$ (although we still require $b \ll r$).
This suggests that the success of the classical line density model to explain apocentre glow in observations may, in fact, be due to the presence of an internal perturber.
In the absence of such a perturber one would expect the apocentre surface brightness enhancement to disappear as the resolution is increased.
This might be one reason for the lack of such a brightness ratio in the ALMA observations of HR\,4796 \citep[see][]{Kennedy18}, despite this disc having been observed as eccentric at much shorter wavelengths \citep[see][]{Olofsson2019}.

\begin{table}
\centering
\caption{Resolution regime for four debris discs that have been interpreted as being eccentric. References for these disc radii ($r$), widths ($\delta r$) and beam sizes can be found in the following sources: 1: \citet{MacGregor17}, 2: \citet{Kennedy18}, 3: \citet{Faramaz19}, 4: \citet{Booth21}. A ``narrow'' disc indicate there exists a beam size where the classical line density could give a good match to the observations. This is not possible for the ``wide'' disc HD\,38206. }
\begin{tabular}{ l | c c c c c r}
\hline
\hline
Source & $r$ & $\delta r$ & Beam & Obs. & Regime & Ref. \\
& [au] & [au] & [au] & & & \\
\hline
Fomalhaut & 143 & 14 & ${\sim}$12 & \textit{ALMA} & ``narrow'' & 1 \\
HR\,4796 & 79 & 10 & ${\sim}$12 & \textit{ALMA} & ``narrow'' & 2 \\
HD\,202628 & 154 & 22 & ${\sim}$21 & \textit{ALMA} & ``narrow'' & 3 \\
HD\,38206 & 180 & 140 & ${\sim}$60 & \textit{ALMA} & ``wide'' & 4 \\
\hline
\end{tabular}
\label{tab:discRegimes}
\end{table}

\subsection{Disc Integrated Flux Enhancements: inconclusive determinant of eccentricity profiles}
\label{sec:discFLUX}

\begin{figure}
    \includegraphics[width=\linewidth]{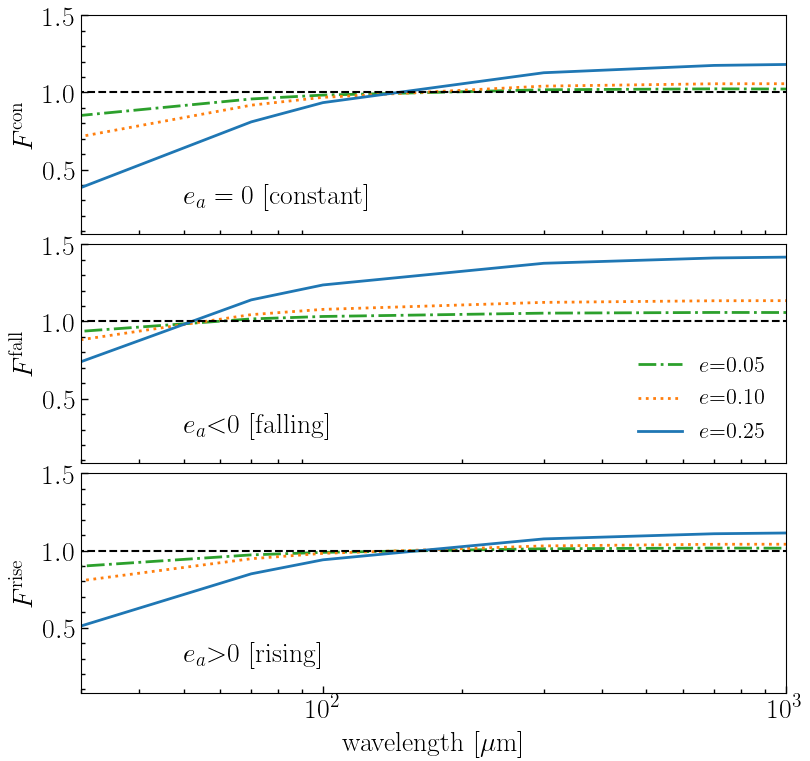}
    \caption{Apocentre-pericentre integrated flux enhancements for discs with $e(a_0){=}0.05$, $0.10$ and $0.25$. Top: constant eccentricity discs, middle: falling eccentricity discs, bottom: rising eccentricity discs.}
    \label{fig:IntFluxRatios}
\end{figure}

\begin{figure}
    \includegraphics[width=\linewidth]{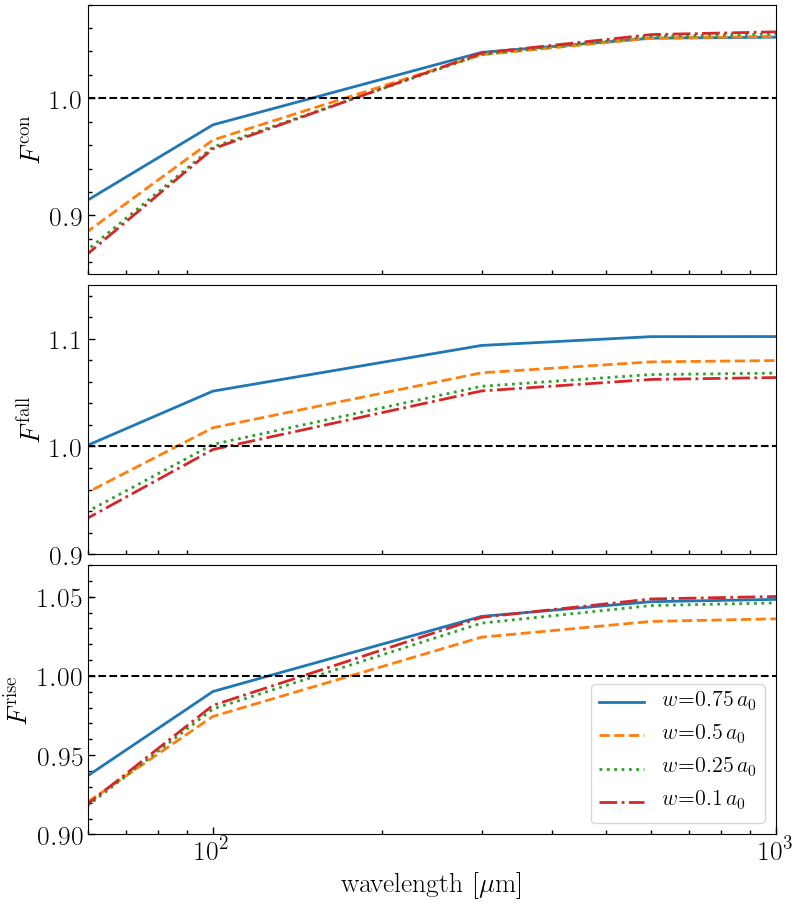}
    \caption{Plots to demonstrate the apocentre-pericentre total flux enhancements, for all three disc models (where top: e=constant, middle: e=falling, and bottom: e=rising), for a fixed $e(a_0){=}0.10$, and varying width.}
    \label{fig:expWidth}
\end{figure}

We considered the integrated flux ratios of the discs in the same regimes as discussed in $\S$\ref{sec:discBrightness}, by integrating all emission ${>}10$\% the peak disc surface brightness (to remove emission far from the star which lowers the overall measured flux ratio) though we note here that given the multitude of ways discs can be integrated (e.g., based on different contour lines) choosing different integration boundaries can lead to a significant variation in measurements. 
We find these vary as a function of wavelength (for the three considered eccentricities of $e_0=0.05$, $0.10$ and $0.25$, also referenced at disc semi-major axis) as shown in Fig.~\ref{fig:IntFluxRatios}. 
Whilst the wavelength at which the apocentre integrated flux dominates is not the same (i.e., for either the same scenario with a different eccentricity, or simply in the different scenarios) in all cases it can be seen that the integrated flux is dominated by the apocentre side of the disc in the long-wavelength regime (i.e., in the sub-mm wavelengths, in accordance with our results from $\S$\ref{sec:Theory1}). 
Further, we extended our models to investigate the disc width. 
We present Fig.\ref{fig:expWidth}, which demonstrates that for discs ranging between narrow and broad (with fractional widths ranging from $w=10\%\,r_0$ to $w=75\%\,r_0$) the total flux integral in the long-wavelength regime remains apocentre-dominated. 
This combination of these leads us to an important observation: at long wavelengths, the integrated flux metric is less conclusive in determining the eccentricity profiles than the surface brightness enhancement (as discussed in $\S$\ref{sec:discBrightness}) and thus less able to distinguish between the origin of the eccentricity distribution. 
We note here however that the measurement of an integrated flux enhancement in the pericentre of a debris disc might therefore provide hints that the underlying structure is not dominated by dust in eccentric orbits.  

\subsection{Other complications?}
There are several effects, not considered here, which can alter the observed morphology of a debris disc. For example, our work has not considered the influence of particles having free eccentricity, whereas this can influence dust density distributions and thus their long-wavelength behaviour. For a more complete investigation of how particle free eccentricities affect debris disc morphologies, we refer the reader to \citet{Kennedy20}.

Since we only considered the effects of face-on discs, additional complications in the dust distributions are introduced when discs become inclined, and by incorporating the effects of finite disc thickness. We intend to follow up on this in detail in future work.


\section{Conclusions}
\label{sec:conclusions}
In this work we have explored and developed the theory of eccentric debris disc morphology using a mixture of analytical and numerical radiative transfer modelling.
We have focused on the limited case of a face on disc with negligible free eccentricity, but with variable eccentricity profiles (i.e., with either positive, flat or negative eccentricity gradients). 
We have shown that, even with these rather limiting assumptions, eccentric debris discs have a diverse set of  morphologies which can complicate attempts to obtain disc (and wider planetary system) properties from observations. In summary our conclusions are:

\begin{enumerate}
 \item A well resolved constant $e$ disc has no apocentre glow at any wavelength.
 \item Eccentricity gradients (and disc twists) cause surface density variations in debris discs. At long wavelengths this causes an enhancement of pericentre glow when $e$ increases outwards, and can result in apocentre glow when $e$ decreases outwards.
 \item A poorly resolved, narrow, constant $e$ disc can have apocentre glow at long wavelengths, but this will switch to pericentre glow at higher resolutions (or for wider discs).
 \item For diagnosing the eccentricity distribution of a debris disc, the relative surface brightness of the apses offers a more conclusive measurement than its integrated flux.
\end{enumerate}

We find that the classical approach of interpreting eccentric debris discs using line densities is only valid under an extremely limited set of circumstances, which are unlikely to be met as debris disc observations become increasingly better resolved.

Finally, we note that 3D disc structure (resulting from disc inclinations and vertical structure) complicate matters significantly, and can result in discs with multiple surface brightness maxima. We aim to explore these 3-dimensional effects in detail in future work.

\section*{Acknowledgements}
We would like to thank Mark Wyatt (MW), Sebastian Marino (SM) and Guillaume Laibe for reviewing the pre-submission manuscript. Additionally we thank MW for useful discussions about the effects of resolution, and SM for sharing disc modelling code. We thank the anonymous reviewer for their comments which improved the quality of this work.

E. Lynch would like to thank the Science and Technologies Facilities Council (STFC) for funding this work through a STFC studentship, and the European Research Council (ERC). This research was supported by STFC through the grant ST/P000673/1 and the ERC through the CoG project PODCAST No 864965. This project has received funding from the European Union’s Horizon 2020 research and innovation program under the Marie Skłodowska-Curie grant agreement No 823823. 

J. Lovell would like to thank the STFC for funding this work through a postgraduate studentship.

\section*{Data availability}

The data underlying this article will be shared on reasonable request to the corresponding author.



\bibliographystyle{mnras}
\bibliography{REFS} 


\appendix
\onecolumn

\section{Analysis of the Long Wavelength Limit} \label{long wav limit}
\begin{figure}
\includegraphics[width=\linewidth]{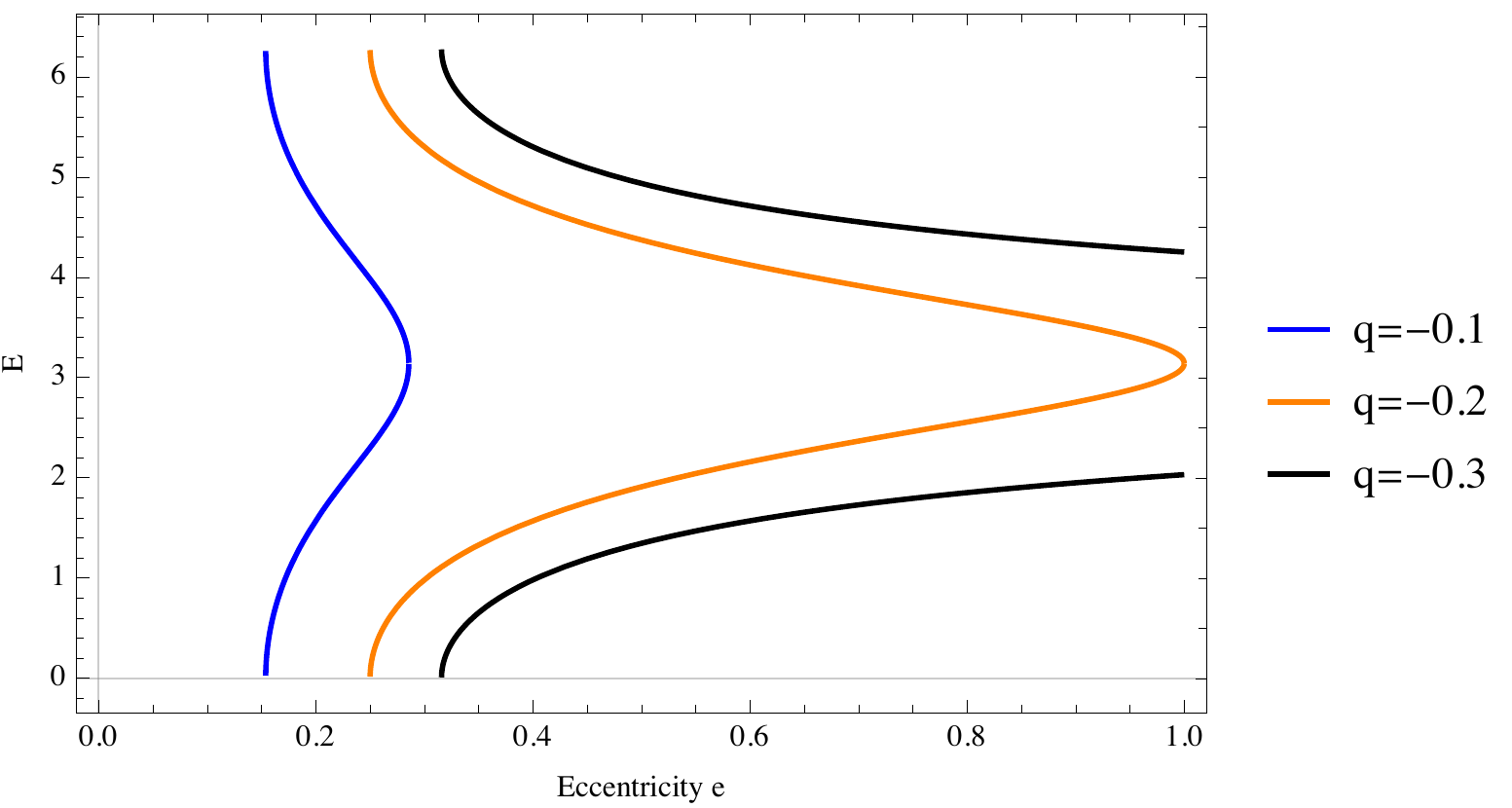}
\caption{Locations of the off apse roots as a function of eccentricity for two different values of $q$. For low (i.e. realistic) values of $q$ the roots only exist for a narrow range of eccentricity. These roots merge with either apocentre or pericentre when the solution transitions from having 4-roots to having 2-roots.}
\label{root location}
\end{figure}

In this Appendix we further explore the long wavelength limit of the 2D theory. The long wavelength (or high temperature) limit ($\lambda \gg \lambda_{*}$) of the non-dimensional surface brightness (see Equation \ref{surf bright eq}) is
\begin{equation}
\lim_{\lambda/\lambda_{*} \rightarrow \infty} \mathcal{I} \propto \frac{1}{(1 - q \cos (E-\alpha)) \sqrt{1 - e \cos E}} \quad .
\label{long wavelength s bright}
\end{equation}
Unlike the short wavelength limit ($\lambda \ll \lambda_{*}$) where the brightest point is always located at pericentre, the location of the surface brightness maxima in Equation \ref{long wavelength s bright} is influenced by both the surface density and the temperature around the orbit. The effects of surface density and temperature can compete with each other, making the location of the surface brightness maxima sensitive to the details of the disc geometry at long wavelengths. 
In general the azimuthal extrema in the surface brightness of an \textit{untwisted} disc are located at
\begin{equation}
\lim_{\lambda/\lambda_{*} \rightarrow \infty} \frac{\partial \mathcal{I}}{\partial E} = -\frac{q \sin E (1 - e \cos E) + \frac{1}{2} e \sin E (1 - q \cos E)}{(1 - q \cos E)^2 (1 - e \cos E)^{3/2}} = 0 ,
\label{extrema locations}
\end{equation}
where we now adopt the convention that $q$ is a signed quantity, with a change of sign corresponding to a rotation of the line of $\alpha$ by $\pi$. Equation \ref{extrema locations} has 4 roots, 2 corresponding to the apses ($E = 0$, $E = \pi$) and 2 additional roots which need not correspond to pericentre/apocentre in general,
\begin{equation}
\cos E = \frac{2 q + e}{3 q e} \quad .
\label{nonaligned roots}
\end{equation}
In order for these roots to exist we require
\begin{equation}
\Biggl| \frac{2 q + e}{3 q e} \Biggr| < 1 \quad .
\label{quartic roots condition}
\end{equation}
The location of these off apse roots is shown in Figure \ref{root location}, for representative values of $q$.
For the root at pericentre we have
\begin{equation}
\lim_{\lambda/\lambda_{*} \rightarrow \infty} \sgn \left(\frac{\partial^2 \mathcal{I}}{\partial E^2}\right) \Biggl |_{E=0} = \sgn \left(-\frac{e}{2}  + \frac{3}{2} e q - q\right) \quad,
\end{equation}
while for apocentre,
\begin{equation}
\lim_{\lambda/\lambda_{*} \rightarrow \infty} \sgn \left(\frac{\partial^2 \mathcal{I}}{\partial E^2}\right) \Biggl |_{E=\pi} = \sgn \left(\frac{e}{2}  +  \frac{3}{2} e q + q \right) \quad .
\end{equation}
Thus we have a maxima at pericentre if
\begin{equation}
\frac{3}{2} e q < q + \frac{e}{2} \quad,
\end{equation}
and a maxima at apocentre if
\begin{equation}
-\frac{3}{2} e q > q + \frac{e}{2} \quad .
\end{equation}
This leaves the interesting possibility that we can simultaneously have a \textit{maxima} at apocentre and pericentre when these inequalities hold simultaneously, i.e.,
\begin{equation}
\frac{3}{2} e q < q + \frac{e}{2} < -\frac{3}{2} e q \quad.
\label{double maxima cond}
\end{equation}
Further, one could in principle have \textit{minima} at both apocentre and pericentre if the following inequality holds 
\begin{equation}
-\frac{3}{2} e q < q + \frac{e}{2} < \frac{3}{2} e q .
\end{equation}
It is straightforward to show that for $q<0$ the first inequality cannot be satisfied. For $q>0$ we consider the second inequality, i.e., 
\begin{equation}
q < \frac{1}{2} e (3 q - 1).
\end{equation}
When $0<q<1/3$ this inequality cannot be satisfied as the right hand side is negative. For $1/3 < q < 1$ we can rearrange the inequality to find
\begin{equation}
 e > \frac{2 q}{3 q - 1} \ge 1 .
\end{equation}
Therefore, the inequality cannot be satisfied for bound, non-overlapping, orbits. As such, the case where both apocentre and pericentre are minima has no physical relevance. We note that Equation \ref{double maxima cond} is the same as the condition for the existence of the off apse roots (Equation \ref{quartic roots condition}). Thus if Equation \ref{quartic roots condition} is violated there only exists extrema at apocentre and pericentre, one of which is a maxima, the other a minima. If Equation \ref{quartic roots condition} is satisfied two additional extrema appear, changing the behaviour of the disc such that both apocentre and pericentre are \textit{maxima}.

The sign of $\frac{\partial^2 \mathcal{I}}{\partial E^2}$ for the off apse root is equal to $\sgn(-q)$, which confirms that these are minima for $q<0$. That these roots do not exist for $q>0$ (where they would be maxima) is straightforward to understand as for $q>0$ there is no competition between surface density and temperature, and so the disc is always brightest at pericentre.

\section{Convolution with a narrow beam} \label{beam convolution explicit}
The observed surface brightnesses can be obtained by convolving the disc surface brightness with the beam,
\begin{equation}
I_{\rm obs} = I * B .
\end{equation}
Without loss of generality, we assume that the major axis of the beam is aligned with the $x$ axis, then the beam is given by the Gaussian
\begin{equation}
B = \frac{1}{2 \pi b_2 b_2} \exp \left( -\frac{x^2}{2 b_1^2} -\frac{y^2}{2 b_2^2} \right) ,
\end{equation}
where $b_1 > b_2$. It is always possible to write the disc mass per unit semi-major axis in the form
\begin{equation}
M_a = \frac{\sqrt{2 \pi}}{w} a \sigma_0(a) \exp \left(-\frac{(a - a_0)^2}{2 w^2} \right) ,
\label{ma appendix}
\end{equation}
where $\sigma_0$ is arbitrary and $w$ is a representative disc length scale (typically the disc width). Assume that $b_2 \ll a_0$ is small enough such that $T(a,E)$, $\sigma_0 (a)$, $e (a)$ and $\varpi (a)$ are all approximately constant within the beam. In the long wavelength limit, $I \approxprop T \Sigma$, and the observed surface brightness is given by
\begin{equation}
I_{\rm obs} \approxprop \frac{1}{(2 \pi)^{3/2} b_1 b_2 w} \iint d x^{\prime} d y^{\prime} \frac{T(a^{\prime},E^{\prime}) \sigma_0(a^{\prime})}{j(a^{\prime},E^{\prime})} \exp \left( - \frac{(a^{\prime} - a_0)^2}{2 w^2} - \frac{(x - x^{\prime})^2}{2 b_1^2} - \frac{(y - y^{\prime})^2}{2 b_2^2} \right)
\end{equation}
We adopting local coordinates $(\chi, \eta)$ such that
\begin{equation}
    a = a_0 + \chi, \quad E = \eta ,
\end{equation}
which are similar to the coordinates introduced in \citet{Ogilvie14}, except based on the $(a,E)$ coordinate system. The associated area element can be obtained by evaluating geometrical properties at $(a,E)$,
\begin{equation}
d x d y = J(a,E) (1 - e \cos E) d \chi d \eta .
\end{equation}
By expanding $x^{\prime}$ and $y^{\prime}$ about $x$ and $y$ we can rewrite the Gaussian beam in terms of the new coordinates, since
\begin{equation}
(x - x^{\prime})^2 + (y - y^{\prime})^2 \approx \gamma_{a a} (\chi - \chi^{\prime})^2 + 2 \gamma_{a E} (\chi - \chi^{\prime}) (\eta - \eta^{\prime})+ \gamma_{E E} (\eta - \eta^{\prime})^2 ,
\end{equation}
where $\gamma_{i j}$ are the components of a `stretched' orbital coordinate system, evaluated at $(a,E)$, and are given by 
\begin{equation}
 \gamma_{a a} = x_a^2 + \frac{b_1^2}{b_2^2} y_a^2 , \quad \gamma_{a E} = x_a x_e + \frac{b_1^2}{b_2^2} y_a y_E , \quad \gamma_{E E} = x_E^2 + \frac{b_1^2}{b_2^2} y_E^2 .
\end{equation}
These equal the metric tensor of the $(a,E)$ orbital coordinate system when $b_2 = b_1$. The observed surface brightness is then given by
\begin{align}
\begin{split}
I_{\rm obs} &\approxprop \frac{1}{(2 \pi)^{3/2} b_1 b_2 w} \iint d \chi^{\prime} d \eta^{\prime} a^{\prime} T(a^{\prime},\eta^{\prime})  (1 - e^{\prime} \cos \eta^{\prime}) \sigma_0 (a^{\prime}) \\
&\times \exp \left[ - \frac{(a - a_0)^2}{2 w^2} - \frac{\gamma_{a a} (\chi - \chi^{\prime})^2 + 2 \gamma_{a E} (\chi - \chi^{\prime}) (\eta - \eta^{\prime})+ \gamma_{E E} (\eta - \eta^{\prime})^2}{2 b_1^2} \right] .
\end{split}
\end{align}
Approximating the integration over the angular variable, $\eta^{\prime}$, using the Laplace method, we obtain
\begin{equation}
I_{\rm obs} \approxprop \frac{1}{(2 \pi) b_2 w} \int d \chi \frac{a^{\prime} T (a^{\prime},E) \sigma_0 (a^{\prime})}{\sqrt{\gamma_{E E}}} (1 - e^{\prime} \cos E) \exp \left[ - \frac{(\chi^{\prime})^2}{2 w^2} - \frac{\mathcal{J}^{2}}{2 b^2 \gamma_{EE}} (\chi - \chi^{\prime})^2 \right] ,
\end{equation}
where we have introduced $\mathcal{J}$ the Jacobian determinant of the stretched coordinate system, with $\mathcal{J}^2 = \gamma_{a a} \gamma_{E E} - \gamma_{a E}^2$ and have approximated $\eta \approx E$ except where it appears in the exponential.
Similarly the integration over $\chi$ can be approximated with Laplace's method. The maximum of the integrand is located at
\begin{equation}
\chi^{\prime} = \left[1 + \frac{b^2 (1 - e_0^2 \cos^2 E)}{w^2 j^2 (a_0,E) (1 - e_0 \cos E)^2} \right]^{-1} \chi .
\end{equation}
The integral is then approximated by,
\begin{equation}
I_{\rm obs} \approxprop \frac{b_1}{\sqrt{2 \pi} w b_2 \mathcal{J}} a T(a,E) \sigma_0(a) (1 - e \cos E) \left(1 + \frac{b_1^2 \gamma_{EE}}{w^2 \mathcal{J}^2} \right)^{-1/2} \exp \left[ - \frac{(a - a_0)^2}{2 w^2} \left(1 + \frac{b_1^2 \gamma_{E E}}{w^2 \mathcal{J}^2} \right)^{-1}\right] ,
\end{equation}
where $\tilde{e} := e(\tilde{a})$ and, similar to the $\eta$ integral, we have approximated $\chi^{\prime} = \chi$ except where it appears in the exponent.
The geometrical parameters $\mathcal{J}$ and $\gamma_{E E}$ are given by,
\begin{equation}
\mathcal{J} = \frac{b_1}{b_2} a j (1 - e \cos E) ,
\end{equation}
\begin{align}
\begin{split}
 \gamma_{E E} &= a^2 (1 - e^2 \cos^2 E) + \left(\frac{b_1^2}{b_2^2} - 1\right) y_E^2 \\
 &= a^2 (1 - e^2 \cos^2 E) + \left(\frac{b_1^2}{b_2^2} - 1\right) a^2 (\sqrt{1 - e} \cos E \cos \varpi - \sin E \sin \varpi)^2 ,
 \end{split}
\end{align}
and the observed surface brightness simplifies to
\begin{equation}
I_{\rm obs} \approxprop \frac{1}{\sqrt{2 \pi} w j}  T(a,E) \sigma_0(a) \left(1 + \frac{b_2^2 (1 - e^2 \cos^2 E) + (b_1^2 - b_2^2) a^{-2} y_E^2}{w^2 j^2 (1 - e \cos E)^2} \right)^{-1/2} \exp \left[ - \frac{(a - a_0)^2}{2 w^2} \left(1 + \frac{b_2^2 (1 - e^2 \cos^2 E) + (b_1^2 - b_2^2) a^{-2} y_E^2}{w^2  j^2 (1 - e \cos E)^2} \right)^{-1}\right] .
\end{equation}
For a symmetric beam ($b_1=b_2 = b$) we thus arrive at
\begin{equation}
I_{\rm obs} \approxprop \frac{1}{\sqrt{2 \pi} w j}  T(a,E) \sigma_0(a) \left(1 + \frac{b^2 (1 + e \cos E)}{w^2 j^2 (1 - e \cos E)} \right)^{-1/2} \exp \left[ - \frac{(a - a_0)^2}{2 w^2} \left(1 + \frac{b^2 (1 + e \cos E)}{w^2  j^2 (1 - e \cos E)} \right)^{-1}\right] .
\end{equation}

\bsp	
\label{lastpage}
\end{document}